\documentclass[pre,showpacs,superscriptaddress,eqsecnum]{revtex4}

\usepackage{amsmath}
\usepackage{amsfonts}
\usepackage{amssymb}

\usepackage{graphicx}

\begin{document}

\title{Fluctuation-induced pressures in fluids in thermal nonequilibrium steady states}

\author{T. R. Kirkpatrick}
\email{tedkirkp@umd.edu}
\affiliation{Institute for Physical Science and Technology, University of Maryland, College Park, Maryland 20877, USA}
\affiliation{Department of Physics, University of Maryland, College Park, Maryland 20877, USA}

\author{J. M. Ortiz de Z\'arate}
\affiliation{Departamento de F\'{\i}sica Aplicada I, Facultad de F\'{\i}sica, Universidad Complutense, 28040 Madrid, Spain}

\author{J. V. Sengers}
\affiliation{Institute for Physical Science and Technology, University of Maryland, College Park, Maryland 20877, USA}

\date{\today}

\begin{abstract}
Correlations in fluids in nonequilibrium steady states are long ranged. Hence, finite-size effects have important consequences in the nonequilibrium thermodynamics of fluids. One consequence is that nonequilibrium temperature fluctuations induce nonequilibrium Casimir-like pressures proportional to the square of the temperature gradient. Hence, fluctuations cause a breakdown of the concept of local thermal equilibrium. Furthermore, transport coefficients become dependent on boundary conditions and on gravity. Thus nonequilibrium fluctuations affect some traditional concepts in nonequilibrium thermodynamics.
\end{abstract}


\maketitle

\section{INTRODUCTION}

Fluctuation induced forces are common in nature~\cite{KardarGolestanian}. A prototype of such a force is the Casimir force between conducting plates due to quantum fluctuations of the electromagnetic (EM) field~\cite{Casimir49}. Its overall strength is set by Planck's constant, $\hbar $, and the force per unit area, or pressure, is~\cite{MostepanenkoTrunov}
\begin{equation} \label{GrindEQ__1_1_}
p_{{\rm EM}} =-\frac{\pi ^{2} }{240} \frac{\hbar c}{L^{4} } ,
\end{equation}
where $L$ is the distance between the plates, and $c$ the speed of light. Equation \eqref{GrindEQ__1_1_} is valid for distances larger than a microscopic length $L_{0}$ which depends on the actual metal of the conducting plates. The minus sign indicates that the Casimir force is an attractive force for two identical conducting plates. More generally, the electromagnetic Casimir force may be attractive or repulsive for two plates of different materials with appropriately chosen intermediate fluids~\cite{DzyaEtAl,MundayCapasso}.

Other commonly discussed forces are those induced by thermal fluctuations in fluids, where the energy scale is set by $k_{{\rm B}}T$, with $k_{{\rm B}}$ is Boltzmann's constant and $T$ the temperature~\cite{KardarGolestanian}. When the thermal fluctuations are large and long range, they will induce forces detectable and satisfying universal laws at distances large compared to molecular length scales~\cite{GambassiEtAl1}. In analogy to the electromagnetic Casimir force, such fluctuation induced forces in condensed matter are also referred to as Casimir or Casimir-like forces. One important case of this type is the Casimir force induced by critical fluctuations in fluids, originally predicted by Fisher and de Gennes~\cite{FisherDeGennes}. In principle, one should expect Casimir forces induced by density fluctuations in the vicinity of a vapor-liquid critical point or by concentration fluctuations in a liquid mixture near a critical point of mixing~\cite{KrechBook,Krech1}. Practical studies have all been devoted to the Casimir effect due to critical concentration fluctuations in liquid mixtures, because of the experimental convenience of ambient pressure at which the phenomenon can be observed~\cite{GambassiEtAl1,Krech2,GambassiEtAl2}. One finds a scale-dependent force per unit area, to be denoted as a critical Casimir pressure $p_{{\rm c}}$, which is given by~\cite{GambassiEtAl1}
\begin{equation} \label{GrindEQ__1_2_}
p_{{\rm c}} =\frac{k_{{\rm B}} T}{L^{3} }~\Theta\!\left(L/\xi \right),
\end{equation}
where $\Theta$ is a finite-size scaling function with $\xi$ being the (bulk) correlation length of the critical fluctuations. For studying the magnitude of the effect one defines a (critical) Casimir amplitude, $\Delta =\mathop{\lim }\limits_{L/\xi \to 0} \Theta\!\left(L/\xi \right)$, which for the universality class of Ising-like systems may vary from $-0.01$ to $+2$ depending on the boundary conditions~\cite{Krech1,Krech2}. That is, the critical Casimir force can be either repulsive or attractive~\cite{RafaiEtAl}. Note that at larger $L$, $\left|p_{{\rm c}} \right|{\rm >}\left|p_{{\rm EM}} \right|$. That is, the critical fluctuations that cause $p_{{\rm c}}$  are effectively of longer range than the electromagnetic fluctuations that cause $p_{{\rm EM}}$. Casimir-like forces are to be expected in fluids, whenever long-range correlations are present, such as those induced by Goldstein modes in superfluids and in liquid crystals~\cite{KardarGolestanian}.

In addition to the Casimir forces mentioned above, it is also known that velocity fluctuations in equilibrium fluids induce forces on the boundaries that are comparable to Van der Waals forces and that decay as inverse cube of wall separation~\cite{Jones,ChanWhite}. The present article is concerned with Casimir-like forces induced by thermal fluctuations in fluids in thermal nonequilibrium steady states (NESS). Specifically, we shall demonstrate the existence of significant nonequilibrium (NE) Casimir-like forces in fluids in the presence of a temperature gradient $\nabla T$. Long-range correlation phenomena are ubiquitous in fluids as reviewed by Dorfman \textit{et al}.~\cite{DorfmanKirkpatrickSengers}. For fluids in thermodynamic equilibrium, couplings between hydrodynamic modes cause long-range dynamic correlations that are responsible for the presence of algebraic long-time tails in the correlation functions for the transport coefficients~\cite{ErnstHaugeVanLeeuwen0,ErnstHaugeVanLeeuwen,ErnstHaugeVanLeeuwen2,PomeauResibois,KirkpatrickBelitzSengers} and for the divergence of certain transport properties near critical points~\cite{KadanoffSwift,Kawasaki,Sengers2}. In nonequilibrium fluids not only the dynamic correlation functions but also the static correlation functions become long range, yielding a pronounced enhancement of the intensity of thermal nonequilibrium fluctuations. The most studied case is a quiescent fluid in the presence of a uniform temperature gradient, $\nabla T$. Then the nonequilibrium contribution to the intensity of the temperature fluctuations is given by~\cite{DorfmanKirkpatrickSengers,KirkpatrickBelitzSengers}
\begin{equation} \label{GrindEQ__1_3_}
\left\langle \left(\delta T\left({\bf k}\right)\right)^{2} \right\rangle _{{\rm NE}} =\frac{k_{{\rm B}} T}{\rho D_{T} \left(\nu +D_{T} \right)} \frac{\left(k_{\parallel } \nabla T\right)^{2} }{k^{6} } .
\end{equation}
Here the temperature gradient is in the $z$-direction in a fluid bounded by two horizontal plates, located at $z=0,L$, while $k$ is the magnitude of the wave vector ${\bf k}=\{k_x,k_y,k_z\}$ and $k_{\parallel }$ the magnitude of its component ${\bf k}_{\parallel}=\{k_x,k_y\}$ parallel to the plates, \textit{i.e.}, perpendicular to the direction of the temperature gradient.  In this equation $\rho$ is the mass density, $D_{T}$ the thermal diffusivity, and $\nu$ the kinematic viscosity. This result was first predicted by Kirkpatrick \textit{et al}.~\cite{KirkpatrickEtAl}, who extended the mode-coupling theory for dealing with long-time correlations in equilibrium systems~\cite{ErnstHaugeVanLeeuwen0,ErnstHaugeVanLeeuwen,ErnstHaugeVanLeeuwen2,PomeauResibois,KirkpatrickBelitzSengers,KadanoffSwift,Kawasaki} to systems out of equilibrium~\cite{KirkpatrickEtAl,KirkpatrickEtAl4,KirkpatrickEtAl3}. The same result was subsequently shown to follow from a rather straightforward extension of fluctuating hydrodynamics to systems out of equilibrium~\cite{RonisProcaccia,LawSengers,BOOK}. These temperature fluctuations can be measured by Rayleigh-scattering experiments which have accurately confirmed the validity of Eq.~\eqref{GrindEQ__1_3_}~\cite{SegreEtAl1,Mixtures3}.

Equation~\eqref{GrindEQ__1_3_} is valid for wave lengths (\textit{i.e.,} inverse wavenumbers), much smaller than the plate separation $L$, such as those accessible in light-scattering experiments~\cite{SegreEtAl1,Mixtures3}. However, from Eq.~\eqref{GrindEQ__1_3_} we see that, for wave vectors with a non-vanishing component parallel to the plates, the intensity of the NE temperature fluctuations will diverge as $k^{-4}$ when $k\to 0$, leading to large finite-size effects for fluid layers bounded between two surfaces~\cite{Physica,EPJ}. In this paper we show that as a consequence the long-range NE fluctuations induce significant Casimir-like forces in fluids in the presence of a temperature gradient. Specifically, for the scale-dependent NE fluctuation contribution, $p_{{\rm NE}} ({z};L)$, to the pressure, we find
\begin{equation} \label{GrindEQ__1_4_}
p_{{\rm NE}} \left(z;L\right)=\frac{c_{p} k_{{\rm B}} T^{2} \left(\gamma -1\right)}{96\pi D_{T} \left(\nu +D_{T} \right)} \left[1-\frac{1}{\alpha c_{p} } \left(\frac{\partial c_{p} }{\partial T} \right)_{p} +\frac{1}{\alpha ^{2} } \left(\frac{\partial \alpha }{\partial T} \right)_{p} \right] F(\tilde{z})~L\left(\frac{\nabla T}{T} \right)^{2} .
\end{equation}
Here $c_{p}$ is the isobaric specific heat capacity, $\gamma$ the ratio of isobaric and isochoric heat capacities, and $\alpha$ the thermal expansion coefficient. The NE pressure depends on the location in the fluid layer through the function
\begin{equation}\label{E15}
F(\tilde{z}) = 6\tilde{z}(1-\tilde{z}),
\end{equation}
where $\tilde{z}=z/L$. Substituting $F(\tilde{z})=1$ in Eq.~\eqref{E15} yields the average fluctuation-induced NE pressure given in our previous publication~\cite{miPRL2}.

Note that for a fixed value of the temperature gradient, the NE pressure actually grows with increasing $L$. This anomalous behavior is a reflection of the very long-ranged spatial correlations in a fluid in a temperature gradient. We note that the intensity of critical fluctuations diverges only as $k^{-2}$ when $k\to 0$, which means that in real space the critical correlations still decay as a function of the distance $r$~\cite{Fisher64}. However, the divergence of the NE fluctuations as $k^{-4}$ implies that the NE correlations actually increase in real space with increasing distance $r$, so that finite-size effects are always present in nonequilibrium systems. The finite-size effects depend on the actual boundary conditions for the fluctuations at the surfaces of the plates. Equations~\eqref{GrindEQ__1_4_} and~\eqref{E15} correspond to stress-free boundary conditions for which we are able to obtain an explicit analytic expression. We note that a fluctuation-induced NE pressure contribution will always be present at any density and temperature of the fluid state, while the critical Casimir effect is only encountered at a restricted range of temperatures and densities or concentrations close to a critical point. Hence, in contrast to the critical Casimir effect, the NE Casimir effects, and those caused by gauge fluctuations and Goldstone modes, are generic effects existing in entire phases, and are not restricted to special states.

To understand the physical origin of the NE pressures, we note that a temperature gradient can cause normal stresses or pressures if nonequilibrium thermodynamics is extended to include nonlinear effects. Specifically, we consider a nonlinear Onsager-like cross effect causing a nonequilibrium contribution to the pressure induced by a temperature gradient:
\begin{equation} \label{GrindEQ__1_5_}
p_{{\rm NE}} =\kappa _{{\rm NL}} \left(\nabla T\right)^{2}.
\end{equation}
In Eq.~\eqref{GrindEQ__1_5_} $\kappa _{{\rm NL}}$ is a nonlinear kinetic coefficient commonly referred to as a nonlinear Burnett coefficient~\cite{Brey}. It is well known that Burnett coefficients, which go beyond the linear transport coefficients in the ordinary Navier-Stokes equations, do not exist due to the presence of long-time-tail (LTT) effects in the corresponding molecular time-dependent correlation functions~\cite{Brey,PomeauResibois,ErnstDorfman}. Instead, one expects that $\kappa _{{\rm NL}}$ contains a LTT contribution which causes $\kappa _{{\rm NL}}$ to diverge linearly with the system size. To account for this phenomenon we need to consider $\kappa _{{\rm NL}} $ as the sum of a bare molecular contribution $\kappa _{{\rm NL}}^{\left(0\right)}$ associated with short-range correlations and a long-range fluctuating hydrodynamics contribution $L\kappa _{{\rm NL}}^{\left(1\right)}$ diverging as $L\to \infty $:
\begin{equation} \label{GrindEQ__1_6_}
\kappa _{{\rm NL}} =\kappa _{{\rm NL}}^{\left(0\right)} +\kappa _{{\rm NL}}^{\left(1\right)} L.
\end{equation}
Substituting $\kappa _{{\rm NL}} \simeq \kappa _{{\rm NL}}^{\left(0\right)}$ into Eq. \eqref{GrindEQ__1_5_} yields a pressure contribution due to short-range correlations. Since the ratio $\kappa _{{\rm NL}}^{\left(0\right)} /\kappa _{{\rm NL}}^{\left(1\right)} L$ will be of the order of $\sigma /L$, where $\sigma$ is a typical intermolecular distance, this contribution is small and can be neglected. Also, such an effect is one of several at molecular scales, including accommodation of the velocity and kinetic energy of the molecules with the wall, which do not satisfy universal laws and should not be characterized as Casimir effects~\cite{GambassiEtAl1}. However, substitution of $\kappa _{{\rm NL}} \simeq \kappa _{{\rm NL}}^{\left(1\right)} L$ into Eq. \eqref{GrindEQ__1_5_} yields a genuine NE Casimir-like pressure due to long-range correlations that are present at $L\gg\sigma$:
\begin{equation} \label{GrindEQ__1_7_}
p_{{\rm NE}} \left(L\right)=\kappa _{{\rm NL}}^{\left(1\right)} L\left(\nabla T\right)^{2} .
\end{equation}
Hence, from  Eqs.~\eqref{GrindEQ__1_4_} and~\eqref{GrindEQ__1_7_} we conclude that the NE pressure is a direct consequence of the divergence of the corresponding nonlinear Burnett coefficient $\kappa _{{\rm NL}} $.

We shall proceed as follows. First, in Sec.~\ref{S2} we shall derive a relationship between the NE pressure and the NE temperature-temperature correlation function. This relationship can be derived from NE statistical mechanics by expressing the nonequilibrium pressure in terms of the integral over an unequal-time current-current correlation function which can be evaluated by a method previously employed in~\cite{KirkpatrickEtAl,KirkpatrickEtAl4,KirkpatrickEtAl3} for nonequilibrium steady states  (Sec.~\ref{S2A}). We also show that the same relationship can be obtained from a Taylor expansion of the pressure fluctuations (Sec.~\ref{S2B}). The NE temperature-temperature correlation function for a fluid between two horizontal plates is then evaluated in Sec.~\ref{S3}. The NE temperature fluctuations and, hence, the NE pressure depend on the boundary conditions for the fluctuations at the surfaces of the plates. In Section~\ref{S3A} we evaluate the nonequilibrium temperature fluctuations and the resulting NE pressure assuming stress-free boundary conditions, for which we are able to obtain an explicit analytic solution. In Section~\ref{S3B} we consider the alternative case of rigid boundaries for which we derive a Galerkin approximation of the NE pressure. In Sec.~\ref{S4} we present estimates for the magnitude of the NE pressure for some liquids. We find that the NE pressure becomes significant at distances much larger than those at which electromagnetic and critical Casimir forces have been observed. Furthermore, it turns out that the thermal NE fluctuations, in contrast to critical fluctuations, become sufficiently long range to be affected not only by the boundaries, but also by gravity. Gravity effects on the NE fluctuations have been predicted on the basis of NE fluctuating hydrodynamics~\cite{SegreSchmitzSengers,Physica2,miPRE}, and they have been confirmed by light-scattering~\cite{VailatiGiglio1} and shadow-graph experiments~\cite{TakacsEtAl2}. In Sec.~\ref{S5} we show how gravity also affects the NE pressure. For fluid layers heated from above, gravity has only a modest influence on the magnitude of the NE pressure. However, for fluid layers heated from below, we find that the NE pressure diverges as the critical value of the Rayleigh number, associated with the onset of convection, is approached. In Sec.~\ref{S6} we shall consider some consequences of our results for the long-time and long-range behavior of correlation functions associated with linear and non-linear transport coefficients.  Our conclusions are summarized in Sec.~\ref{S7}.

\section{NE PRESSURE AND NE TEMPERATURE FLUCTUATIONS\label{S2}}

\subsection{Statistical-Mechanical Derivation\label{S2A}}

In statistical mechanics the pressure is given by the diagonal element of the microscopic stress tensor averaged over the \textit{N}-particle distribution function, $\rho _{N} $. Here we consider a fluid in a nonequilibrium steady state (NESS) that is close to local equilibrium. We can then decompose $\rho _{N}$ into a local-equilibrium part, $\rho_{{\rm LE}}$, and a part linear in the macroscopic gradients, $\rho_{\nabla}$. The explicit expression for the local-equilibrium part is
\begin{equation} \label{GrindEQ__2_1_}
\rho _{{\rm LE}} =\frac{\exp [y{\rm \star }a]}{{\rm Tr\; exp}[y{\rm \star }a]}
\end{equation}
with $\{a\}=\{\rho ,m\mathbf{v},e\}$ the set of microscopic conserved quantities, $\{y\}=\{\beta \mu - \beta u^{2} /2, \beta {\bf u}, -\beta\}$ the macroscopic conjugate variables, while $y{\rm \star }a=\int d{\bf r}~ y({\bf r})~a\left({\bf r}\right)$ denotes an integration over space. In these expressions, $\rho$ is the mass density, $m{\bf v}$ is the momentum density, $e$ is the energy density, $\beta =1/k_{{\rm B}} T$ is the inverse temperature, $\mu$ is the chemical potential, and ${\bf u}$ is the fluid velocity with magnitude $u$. In a nonequilibrium steady state of a one-component fluid with a temperature gradient, but no velocity gradient, Liouville's equation gives for the gradient part of the \textit{N}-particle distribution function a time-dependent integral of the form \cite{Zubarev}
\begin{equation} \label{GrindEQ__2_2_}
\rho _{\nabla } =-\int _{0}^{\infty }dt{\kern 1pt} \exp \left(-\mathcal{L}t\right)\rho _{{\rm LE}}  \widehat{{\bf J}}_{e} {\rm \star }~\frac{\partial y_{e} }{\partial {\rm x}} .
\end{equation}
Here $\mathcal{L}$ is Liouville's operator, $\widehat{{\bf J}}_{e}$ is the part of the energy current that is orthogonal to the conserved quantities ($\widehat{{\bf J}}_{e} ={\rm P}_{\bot } {\bf J}_{e}$ with ${\bf J}_{e}$ the energy current and ${\rm P}_{\bot }$ a projection operator), and $y_{e} =-\beta$. With $J_{l}$ being the diagonal element of the microscopic stress tensor, the nonequilibrium or gradient part of the pressure tensor can be written as
\begin{equation} \label{GrindEQ__2_3_}
\begin{split}
p_{{\rm NE}} \left({\bf r}\right)&=\left\langle J_{l} \left({\bf r}\right)\right\rangle _{{\rm NE}}\\
&=-\int _{0}^{\infty }dt{\kern 1pt} \left\langle J_{l} \left({\bf r}{\it ,t}\right)\widehat{{\it J}}_{e} \left(0\right)\right\rangle _{{\rm LE}}  {\rm \star }\frac{\partial y_{e} }{\partial {\rm x}}.
\end{split}
\end{equation}
For reference we note that Eq.~\eqref{GrindEQ__2_3_} also implies that the nonequilibrium contribution to the equal-time conserved quantity correlation function is,
\begin{equation} \label{E24}
D_{\alpha\beta}(\mathbf{r}_1,\mathbf{r}_2)=-\int _{0}^{\infty }dt{\kern 1pt} \left\langle a_\alpha(\mathbf{r}_1,t) a_\beta(\mathbf{r}_2,t) \hat{\mathbf{J}}_e(0)\right\rangle _{{\rm LE}}  {\rm \star }\frac{\partial y_{e} }{\partial {\rm x}}.
\end{equation}
Here $\left\langle \right\rangle _{{\rm NE}}$ denotes a nonequilibrium ensemble average and $\left\langle \right\rangle _{{\rm LE}}$ denotes a local-equilibrium ensemble average. Generally, $p_{{\rm NE}} \left({\bf r}\right)$ is a local NE pressure depending on the position ${\bf r=}\left\{x,y,z\right\}$. Equation \eqref{GrindEQ__2_3_} has the structure of a Green-Kubo expression for a transport coefficient, namely, an unequal time current-current correlation function, integrated over all times \textit{t}~\cite{Zwanzig65}. Note, however, that the currents in the integrand of this equation are different, unlike the current-current correlation functions for the usual Navier-Stokes transport coefficients \cite{ErnstHaugeVanLeeuwen}. Hence, the NE pressure originates from a cross Onsager-like effect, \textit{i.e., }a normal stress or pressure is caused by a temperature gradient in accordance with Eq. \eqref{GrindEQ__1_5_}. This point will be further discussed in Sec. VI.

Techniques to evaluate the long-wavelength, or hydrodynamic-mode, contributions to local-equilibrium correlation functions like Eq.~\eqref{GrindEQ__2_3_} have been developed by Kirkpatrick \textit{et al.} \cite{KirkpatrickEtAl,KirkpatrickEtAl4,KirkpatrickEtAl3}, who extended  the methods of Ernst \textit{et al.} \cite{ErnstHaugeVanLeeuwen,ErnstHaugeVanLeeuwen2} to nonequilibrium steady states. The basic idea is as follows. First the local equilibrium average in Eq.~\eqref{GrindEQ__2_3_} is written as two averages. The first average is a constrained LE average where there is a fixed initial current distribution $\{\hat{J}\}$, and where the conserved quantities $\{a\}$ are given specified values $\{A\}$ at $t=0$. The second average is over all values of $\{\hat{J}\}$ and $\{A\}$. The unknown time dependence in Eq.~\eqref{GrindEQ__2_3_} is then transferred into a time-dependent constrained ensemble $F_{JA}(t)$. A local-equilibrium assumption is used, $F_{JA}(t)\simeq F_\text{LE} [y_\text{ss}+\delta{y}(t)]+$corrections that are rapidly decaying or are higher order in the gradients and can be neglected. Here $y=y_\text{ss}+\delta{y}$ are the hydrodynamic fields in the constrained ensemble. In particular, in the formalism the average $\langle\hat{J}_e\rangle_{F_\text{LE}[y_\text{ss}+\delta{y}]}$ appears. Expanding in powers of $\delta{y}\star{a}$ and using that the projected currents are orthogonal to a single conserved quantity, we find that the leading non-zero contributions are the two-mode, or mode-coupling, terms $\sim\delta{y}(t)\delta{y}(t)\star\star\langle{aa}\hat{J}_e\rangle_\text{LE}$. Using all of this and Eq.~\eqref{E24}, we can write Eq.~\eqref{GrindEQ__2_3_} in terms of a $D$ and another mode-coupling amplitude $\sim\langle{J}_l(\mathbf{r})s_0s_0\rangle_\text{LE}$, where $s_0$ is a microscopic entropy fluctuation at zero wave number. The leading contribution in the case of a temperature gradient is
\begin{equation} \label{GrindEQ__2_4_}
p_{{\rm NE}} \left({\bf r}\right)=\frac{1}{2} \left(\frac{\rho }{k_{{\rm B}} c_{p} } \right)^{2} \left\langle J_{l,0} s_{0} s_{0} \right\rangle D_{ss} \left(\mathbf{r},\mathbf{r}\right),
\end{equation}
where $D_{ss}(\mathbf{r},\mathbf{r})$ is the nonequilibrium contribution to the entropy-entropy correlation function with $s$ being the microscopic entropy density. The subscripts 0 of the phase variables in the statistical-mechanical average in Eq. \eqref{GrindEQ__2_4_} indicate that they are all at zero wave number, so that in accordance with Eq. (A.15) in Ref. \cite{ErnstHaugeVanLeeuwen}, including a conversion of entropy per particle to entropy density:
\begin{equation} \label{GrindEQ__2_5_}
\left\langle J_{l,0} s_{0} s_{0} \right\rangle =\rho c_{p} k_{{\rm B}}^{2} T\left(\gamma -1\right)\left[1-\frac{1}{\alpha c_{p} } \left(\frac{\partial c_{p} }{\partial T} \right)_{p} +\frac{1}{\alpha ^{2} } \left(\frac{\partial \alpha }{\partial T} \right)_{p} \right].
\end{equation}
Rather than the nonequilibrium entropy fluctuations we prefer to relate the NE pressure to the nonequilibrium contribution to the temperature-temperature correlation function $D_{TT} \left({\bf r},{\bf r}\right)$ which is related to $D_{ss} \left({\bf r},{\bf r}\right)$ via
\begin{equation} \label{GrindEQ__2_6_}
D_{ss} \left({\bf r},{\bf r}\right)=\left(\frac{c_{p} }{T} \right)^{2} D_{TT} \left({\bf r},\mathbf{r}\right),
\end{equation}
since $\delta{s}=(c_p/T)~\delta{T}$. Substitution of Eqs. \eqref{GrindEQ__2_5_} and \eqref{GrindEQ__2_6_} into Eq. \eqref{GrindEQ__2_4_} yields for the relationship between the NE pressure and the NE temperature fluctuations
\begin{equation} \label{GrindEQ__2_7_}
p_{{\rm NE}} \left(z\right)=\frac{\rho c_{p} \left(\gamma -1\right)}{2T} \left[1-\frac{1}{\alpha c_{p} } \left(\frac{\partial c_{p} }{\partial T} \right)_{p} +\frac{1}{\alpha ^{2} } \left(\frac{\partial \alpha }{\partial T} \right)_{p} \right]D_{TT} \left(z{\rm ,}z\right).
\end{equation}
We note that in the parallel-plate configuration adopted, the NE pressure will only depend on the $z$ coordinate perpendicular to the plates.

\subsection{Taylor Expansion of Pressure Fluctuations\label{S2B}}

An alternative procedure for relating the NE pressure to the NE temperature fluctuations starts by considering a fluctuating pressure as a function of a fluctuating mass density $\rho+\delta\rho$ and a fluctuating energy density $e+\delta{e}$ and writing the pressure in terms of the local mean values $\rho ,e$ and their fluctuations $\delta \rho ,\delta e$:
\begin{equation} \label{GrindEQ__2_8_}
p(\rho +\delta \rho ,e +\delta{e})=p(\rho ,e )+\delta p.
\end{equation}
In the theory of fluctuations it is most simple to consider the pressure as a function of the conserved thermodynamic quantities $\rho$ and $e$ \cite{Onuki97_2}. Then, we apply a Taylor expansion up to terms quadratic in terms of $\delta\rho$ and $\delta{e}$:
\begin{equation} \label{GrindEQ__2_9_}
\begin{array}{l} {p(\rho +\delta \rho ,e+\delta e)=} \\ {p(\rho ,e)+\left(\frac{\partial p}{\partial \rho } \right)_{e} \delta \rho +\left(\frac{\partial p}{\partial e} \right)_{\rho } \delta e+\frac{1}{2} \left[\left(\frac{\partial ^{2} p}{\partial \rho ^{2} } \right)_{e} \left(\delta \rho \right)^{2} +2\left(\frac{\partial ^{2} p}{\partial \rho \partial e} \right)\delta \rho \delta e+\left(\frac{\partial ^{2} p}{\partial e^{2} } \right)_{\rho } \left(\delta e\right)^{2} \right]} \end{array}.
\end{equation}
We know that the NE enhancement of the temperature fluctuations, given by Eq.~\eqref{GrindEQ__1_3_}, is caused by a coupling of entropy or temperature fluctuations with transverse velocity fluctuations with vanishing linear pressure fluctuations~\cite{KirkpatrickEtAl,LawSengers,BOOK}, so that
\begin{equation} \label{GrindEQ__2_10_}
\left(\frac{\partial p}{\partial \rho } \right)_{e} \delta \rho +\left(\frac{\partial p}{\partial e} \right)_{\rho } \delta e=0,
\end{equation}
and
\begin{equation} \label{GrindEQ__2_11_}
\delta \rho =-\rho \alpha \delta T.
\end{equation}
We now substitute Eqs.~\eqref{GrindEQ__2_10_} and~\eqref{GrindEQ__2_11_} into~\eqref{GrindEQ__2_9_} and determine the average NE contribution $p_{{\rm NE}}$ to the equilibrium pressure $p$:
\begin{equation} \label{GrindEQ__2_12_}
p_{{\rm NE}} \left({\bf r}\right)=\frac{\left(n\alpha \right)^{2} }{2} \left[\left(\frac{\partial ^{2} p}{\partial n^{2} } \right)_{e} -2w\left(\frac{\partial ^{2} p}{\partial n\partial e} \right)+w^{2} \left(\frac{\partial ^{2} p}{\partial e^{2} } \right)_{n} \right]\left\langle \left(\delta T\left({\bf r}\right)\right)^{2} \right\rangle _{{\rm NE}}
\end{equation}
with
\begin{equation} \label{GrindEQ__2_13_}
w=\left(\frac{\partial p}{\partial n} \right)_{e} /\left(\frac{\partial p}{\partial e} \right)_{n} .
\end{equation}
We note that only the NE temperature fluctuations $\left\langle \left(\delta T\right)^{2} \right\rangle _{{\rm NE}}$ cause a renormalization of the pressure, since the equilibrium temperature fluctuations are already incorporated in the un-renormalized pressure. Again with the help of some thermodynamic relations~\cite{ErnstHaugeVanLeeuwen}, Eq.~\eqref{GrindEQ__2_12_} can be converted into
\begin{equation} \label{GrindEQ__2_14_}
p_{{\rm NE}} \left(z\right)=\frac{\rho c_{p} \left(\gamma -1\right)}{2T} \left[1-\frac{1}{\alpha c_{p} } \left(\frac{\partial c_{p} }{\partial T} \right)_{p} +\frac{1}{\alpha ^{2} } \left(\frac{\partial \alpha }{\partial T} \right)_{p} \right]\left\langle \left(\delta T\right)^{2} \right\rangle _{{\rm NE}} .
\end{equation}
We thus recover Eq.~\eqref{GrindEQ__2_7_}, since
\begin{equation} \label{GrindEQ__2_15_}
\left\langle \left(\delta T\right)^{2} \right\rangle _{{\rm NE}} =\left\langle \left(\delta T\left(z\right)\delta T\left(z\right)\right)\right\rangle _{{\rm NE}} =D_{TT} \left(z,z\right).
\end{equation}

\section{BOUNDARY CONDITIONS AND NE PRESSURE\label{S3}}

\subsection{Stress-free boundary conditions\label{S3A}}

Nonequilibrium temperature fluctuations in a fluid layer between two horizontal plates have been evaluated in previous publications~\cite{Physica,EPJ}. The long-range NE temperature fluctuations depend on the actual boundary conditions for the fluctuations at the surfaces of the plates. Here we consider stress-free boundary conditions for the vertical component of the velocity and perfectly heat-conducting walls~\cite{Chandra}:
\begin{align}
\delta T&=0 & \text{at}~z&=0,L,\label{GrindEQ__3_1_}\\
\widehat{{\bf n}}\cdot \delta {\bf v}&=0=\pm\delta{v}_z & \text{at}~z&=0,L,\label{GrindEQ__3_2_}\\
(\nabla \times \delta {\bf v})\times \widehat{{\bf n}}&=0 & \text{at}~z&=0,L,\label{GrindEQ__3_3_}
\end{align}
where  $\widehat{{\bf n}}$ is the unit vector normal to the walls. For incompressible fluids condition \eqref{GrindEQ__3_3_} simplifies to \cite{Physica,Chandra}
\begin{align}
\frac{d^{2} v_{z} }{dz^{2} }&=0 & \text{at}~z&=0,L.\label{GrindEQ__3_4_}
\end{align}
The advantage of slip-free boundary conditions, as opposed to rigid boundaries, is that an explicit analytic solution can be obtained for the NE temperature fluctuations.

Instead of the temperature fluctuations directly, previous publications~\cite{Physica,EPJ} evaluated a nonequilibrium structure factor $S_{{\rm NE}} \left(k_{\parallel } ,z,z'\right)$, which is related to the nonequilibrium spatial correlation function of equal-time temperature fluctuations as \cite{BOOK,Physica}
\begin{equation} \label{GrindEQ__3_5_}
\left\langle \delta T\left({\bf r}\right)\delta T\left({\bf r}{\rm '}\right)\right\rangle _{{\rm NE}} =\frac{1}{\alpha ^{2} \rho ^{2} } \frac{1}{4\pi ^{2} } \iint d\mathbf{k}_{\parallel }~  \exp\!\left[{\rm i}\mathbf{k}_\parallel\cdot \left(\mathbf{x}-\mathbf{x}^\prime\right)\right]~S_{{\rm NE}} \left(k_{\parallel } ,z,z'\right),
\end{equation}
where ${\bf x}$ and ${\bf x}{\rm '}$ are the projections of ${\bf r}$ and ${\bf r}{\rm '}$ onto the plane parallel to the boundaries at $z=0,L$. For the stress-free boundary conditions \eqref{GrindEQ__3_1_}-\eqref{GrindEQ__3_4_}, this nonequilibrium structure factor is given by
\begin{equation}
S_{{\rm NE}} \left(k_{\parallel } ,z,z'\right)=\frac{\rho ^{2} \kappa _{T} c_{p} k_{{\rm B}} }{D_{T} \left(\nu +D_{T} \right)} \frac{\left(\gamma -1\right)}{\gamma } \left(\nabla T\right)^{2} \widetilde{S}_{{\rm NE}} \left(k_{\parallel } ,z,z'\right)\label{GrindEQ__3_6_}
\end{equation}
with
\begin{equation} \label{GrindEQ__3_7_}
\widetilde{S}_{{\rm NE}} \left(k_{\parallel } ,z,z'\right)=2L^{3} \sum _{1}^{\infty }\frac{k_{\parallel }^{2} L^{2} \sin \left(N\pi z/L\right)\sin \left(N\pi z'/L\right)}{\left(k_{\parallel }^{2} L^{2} +N^{2} \pi ^{2} \right)^{3} }
\end{equation}
in accordance with Eqs.~(20) and~(21) in~\cite{Physica}. If we substitute Eq.~\eqref{GrindEQ__3_6_} into Eq. \eqref{GrindEQ__3_5_} and use the relation, $\rho \kappa _{T} c_{p} =\alpha ^{2} T\gamma /\left(\gamma -1\right)$, where $\kappa _{T}$ is the isothermal compressibility, we obtain
\begin{equation} \label{GrindEQ__3_8_}
\left\langle \delta T\left({\bf r}\right)\delta T\left({\bf r}{\rm '}\right)\right\rangle _{{\rm NE}} =\frac{k_{{\rm B}} T\left(\nabla T\right)^{2} }{\rho D_{T} \left(\nu +D_{T} \right)} \frac{1}{4\pi ^{2} } \iint d\mathbf{k}_{\parallel }~  \exp\!\left[{\rm i}\mathbf{k}_\parallel\cdot \left(\mathbf{x}-\mathbf{x}^\prime\right)\right]~\widetilde{S}_{{\rm NE}} \left(k_{\parallel } ,z,z'\right).
\end{equation}
To obtain the NE pressure from Eq. \eqref{GrindEQ__2_14_}, we now need the consider the nonequilibrium part of the temperature fluctuations autocorrelation function
\begin{equation} \label{GrindEQ__3_9_}
\left\langle \delta T^2\right\rangle _{{\rm NE}} =\frac{k_{{\rm B}} T\left(\nabla T\right)^{2} }{\rho D_{T} \left(\nu +D_{T} \right)} \frac{1}{4\pi ^{2} } \iint d{\bf k}_{\parallel }  \widetilde{S}_{{\rm NE}} \left(k_{\parallel } ,z,z\right).
\end{equation}
If we substitute Eq. \eqref{GrindEQ__3_7_} into \eqref{GrindEQ__3_9_} with $z'=z$ and introduce dimensionless variables $\widetilde{k}_\parallel=k_{\parallel } L$ and $\widetilde{z}=z/L$, we obtain
\begin{equation} \label{GrindEQ__3_10_}
\left\langle \delta T^2\right\rangle _{{\rm NE}} =\frac{k_{{\rm B}} TL\left(\nabla T\right)^{2} }{48\pi \rho D_{T} \left(\nu +D_{T} \right)} F\left(\widetilde{z}\right)
\end{equation}
with
\begin{equation} \label{GrindEQ__3_11_}
F\left(\widetilde{z}\right)=48 \int _{0}^{\infty }d\widetilde{k} \sum _{1}^{\infty }\frac{\widetilde{k}^{3} \sin ^{2} \left(N\pi \widetilde{z}\right)}{\left(\widetilde{k}^{2} +N^{2} \pi ^{2} \right)^{3} }  =6  \widetilde{z}\left(1-\widetilde{z}\right).
\end{equation}
Next we substitute Eq.~\eqref{GrindEQ__3_10_} into Eq.~\eqref{GrindEQ__2_14_} so as to obtain for the NE pressure induced by a temperature gradient
\begin{equation} \label{GrindEQ__3_12_}
p_{{\rm NE}} \left(z;L\right)=\frac{c_{p} k_{{\rm B}} T^{2} \left(\gamma -1\right)}{96\pi D_{T} \left(\nu +D_{T} \right)} \left[1-\frac{1}{\alpha c_{p} } \left(\frac{\partial c_{p} }{\partial T} \right)_{p} +\frac{1}{\alpha ^{2} } \left(\frac{\partial \alpha }{\partial T} \right)_{p} \right]F\left(\widetilde{z}\right)L\left(\frac{\nabla T}{T} \right)^{2} .
\end{equation}
We note that the NE pressure depends on the vertical location $z$ and also depends (parametrically) on the distance $L$ between the two horizontal plates. From Eq.~\eqref{GrindEQ__3_12_} the average NE pressure induced in the fluid layer can be easily calculated, resulting in~\cite{miPRL2}
\begin{equation} \label{GrindEQ__3_13_}
p_{{\rm NE}} \left(L\right)=\left\langle p_{{\rm NE}} \left(z;L\right)\right\rangle _{z} =\frac{c_{p} k_{{\rm B}} T^{2} \left(\gamma -1\right)}{96\pi D_{T} \left(\nu +D_{T} \right)} \left[1-\frac{1}{\alpha c_{p} } \left(\frac{\partial c_{p} }{\partial T} \right)_{p} +\frac{1}{\alpha ^{2} } \left(\frac{\partial \alpha }{\partial T} \right)_{p} \right]L\left(\frac{\nabla T}{T} \right)^{2} .
\end{equation}

\begin{figure}[b]
\begin{center}
\includegraphics[width=0.6\columnwidth]{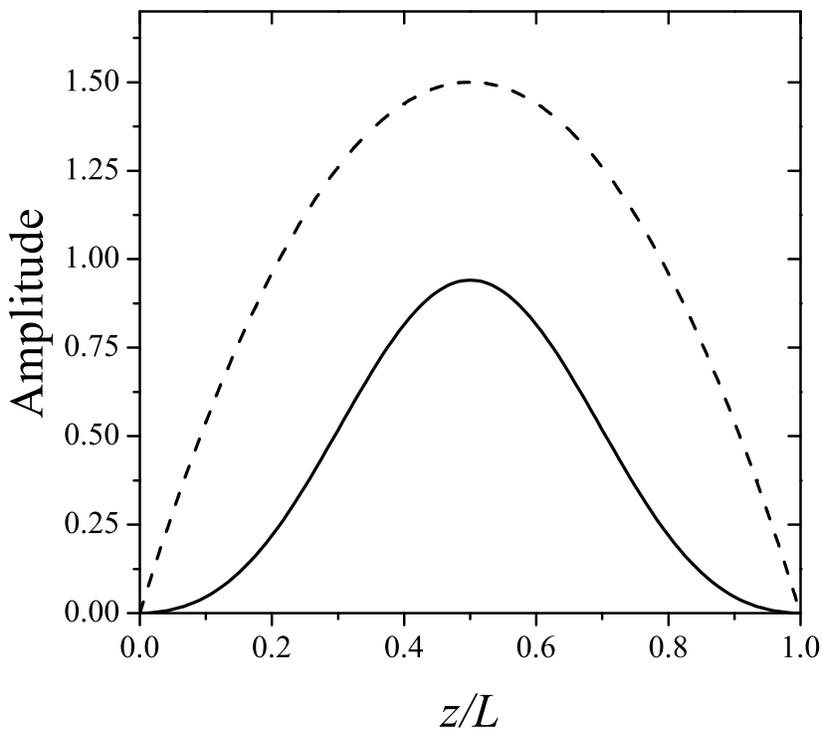}
\end{center}
\caption{Amplitude of NE pressure as a function of $\widetilde{z}$. Dashed curve: the function $F\left(\widetilde{z}\right)$ in Eq. \eqref{GrindEQ__3_12_} for the case of stress-free boundary conditions. Solid curve: the function $\overline{F}\left(\widetilde{z}\right)$ evaluated numerically for the case of rigid boundaries in a Galerkin approximation with ${\rm Pr=6}$.}
\label{F1}
\end{figure}

\subsection{Rigid boundaries\label{S3B}}

While the stress-free boundary conditions are convenient for obtaining the simple and exact expression~\eqref{GrindEQ__3_12_} for the NE pressure, a fluid bounded by two free surfaces may not be a realistic representation of some experimental situations~\cite{Chandra}. For a fluid layer between two rigid walls one commonly imposes no-slip boundary conditions for the local velocity. That is, Eq.~\eqref{GrindEQ__3_4_} is replaced by~\cite{EPJ,Chandra}
\begin{align}
\pm\frac{dv_{z} }{dz} &=0 & \text{at}~z&=0,L,\label{GrindEQ__3_14_}
\end{align}
but we can continue to assume perfectly heat-conducting walls implied by condition~\eqref{GrindEQ__3_1_}. For rigid boundary conditions it is not possible to obtain an explicit analytic expression. Instead we adopt here a Galerkin approximation for the temperature fluctuations obtained elsewhere~\cite{EPJ}. We then find that Eq.~\eqref{GrindEQ__3_7_} for $\tilde{S}_\text{NE}(k_\parallel,z,z^\prime)$ needs to be replaced by Eq.~(26) in~\cite{EPJ}. If we then follow the same procedure as in Subsection~\ref{S3A}, we obtain for the NE-pressure in a fluid layer between two rigid boundaries an expression similar to Eq.~\eqref{GrindEQ__3_12_} before, but where the function $F(\tilde{z})$ for free boundaries is replaced by an equivalent function $\overline{F}(\tilde{z})$ for rigid boundaries. It turns out that for rigid boundaries the amplitude function $\overline{F}(\tilde{z})$ also depends on the Prandtl number ${\rm Pr}=\nu/D_{T}$, unlike $F(\tilde{z})$ for free boundaries.  The explicit expression for the amplitude function $\overline{F}$ can be readily evaluated on the basis of the Galerkin approximation of~\cite{EPJ}, but it results in a somewhat complicated expression. Hence, we prefer here to discuss the function $\overline{F}(\tilde{z})$ graphically.

In Fig.~\ref{F1} we show the functions $F\left(\widetilde{z}\right)$ and $\overline{F}\left(\widetilde{z}\right)$ as a function of the location $\widetilde{z}$ in the fluid layer between the plates. The major difference is that in the case of stress-free boundary condition the slope $dp_{{\rm NE}} \left(z;L\right)/dz$ has a finite limit at $z=0,L$, while in the case of rigid boundary conditions the slope $dp_{{\rm NE}} \left(z;L\right)/dz$ vanishes at the boundaries. We also observe in Fig.~\ref{F1} that the amplitude function is somewhat smaller in the case of rigid boundaries, the average through the layer being, roughly, 40\% less compared to the case of free boundaries. For the average NE pressure we find
\begin{equation} \label{GrindEQ__3_17_}
\begin{split}
p_{{\rm NE}} \left(L\right)&=\left\langle p_{{\rm NE}} \left(z;L\right)\right\rangle _{z}\\
&=\frac{c_{p} k_{{\rm B}} T^{2} \left(\gamma -1\right)}{96\pi D_{T} \left(\nu +D_{T} \right)}\left\langle \overline{F}\left(\widetilde{z}\right) \right\rangle _{z} \left[1-\frac{1}{\alpha c_{p} } \left(\frac{\partial c_{p} }{\partial T} \right)_{p} +\frac{1}{\alpha ^{2} } \left(\frac{\partial \alpha }{\partial T} \right)_{p} \right]L\left(\frac{\nabla T}{T} \right)^{2}.
\end{split}
\end{equation}
where the average amplitude $\left\langle \overline{F}\left(\widetilde{z}\right) \right\rangle _{z}$ over the height of the layer is now a function of the Prandtl number, while for the case of free boundaries it was a constant. An explicit expression for $\left\langle \overline{F}\left(\widetilde{z}\right) \right\rangle _{z}$ as a function of Pr can be obtained from the expressions of~\cite{EPJ}. We have found that for $Pr\gtrsim 1$ the average $\left\langle \overline{F}\left(\widetilde{z}\right) \right\rangle _{z}$  is well represented (within a few percent) by the first two terms of its asymptotic expansion in powers of $\text{Pr}^{-1}$, namely:
\begin{equation}\label{E316}
\left\langle \overline{F}\left(\widetilde{z}\right) \right\rangle _{z} \simeq  A + \frac{B}{\text{Pr}} + \mathcal{O}\left(\frac{1}{\text{Pr}^2}\right)
\end{equation}
with
\begin{align*}
A&\simeq 0.40112, & B&=A-\frac{1}{2\sqrt{10}} \left[\pi-2 \arctan\left( \sqrt{\frac{2}{5}}\right)  \right].
\end{align*}
We emphasize that Eq.~\eqref{E316} for rigid boundary conditions is obtained on the basis of a zeroth-order Galerkin approximation~\cite{EPJ}.

\section{ESTIMATED NE PRESSURES\label{S4} }

For a given value of the temperature gradient $\nabla T$, the NE pressure increases with the distance $L$ between the plates in accordance with Eqs.~\eqref{GrindEQ__3_13_} and~\eqref{GrindEQ__3_17_}. Alternatively, one may study the NE pressure at a given temperature difference $\Delta T=L\nabla T$. For the case of free boundaries, from Eq.~\eqref{GrindEQ__3_13_} one then obtains:
\begin{equation} \label{GrindEQ__4_1_}
\begin{split}
p_{{\rm NE}}(L)& =\langle p_{{\rm NE}}(\tilde{z};L)\rangle_{\tilde{z}} \\
& =\frac{c_{p} k_{{\rm B}} T^{2} \left(\gamma -1\right)}{96\pi D_{T} \left(\nu +D_{T} \right)} \left[1-\frac{1}{\alpha c_{p} } \left(\frac{\partial c_{p} }{\partial T} \right)_{p} +\frac{1}{\alpha ^{2} } \left(\frac{\partial \alpha }{\partial T} \right)_{p} \right]\frac{1}{L} \left(\frac{\Delta T}{T} \right)^{2},
\end{split}
\end{equation}
while for the case of rigid boundaries, the average fluctuation-induced NE pressure at a given temperature difference will be equal to the value for free boundaries~\eqref{GrindEQ__4_1_} multiplied by a factor  $\left\langle \overline{F}\left(\widetilde{z}\right) \right\rangle _{z}$ , that depends only on the Prandtl number. Within the zeroth-order Galerkin approximation of~\cite{EPJ}, we obtain for this multiplicative factor the expression~\eqref{E316} given above.

In either case, the $L$ dependence of the NE pressure $p_{{\rm NE}} \left(L\right)\propto L^{-1} $ at a given temperature difference $\Delta T$ may be compared with the EM Casimir pressure $p_{{\rm EM}} \propto L^{-4} $ in accordance with Eq.~\eqref{GrindEQ__1_1_}, and the critical Casimir pressure $p_{{\rm c}} \propto L^{-3} $ in accordance with Eq.~\eqref{GrindEQ__1_2_}. Hence, the fluctuation-induced NE pressures are present over a much larger range of distances $L$ than either $p_{{\rm EM}}$ or $p_{{\rm c}}$. As an illustration, we present in Table~\ref{T1} the estimated NE pressures for liquid water and liquid heptane with $\Delta T=-25$~K at an average temperature of 25$^\circ$C. In this paper we adopt the convention that $\Delta{T}$ is positive when the temperature gradient is in the same direction as gravity (heated from below) and negative when the temperature gradient is opposite to the direction of gravity (heated from above). The values quoted for $p_\text{NE}$ represent the average value of the NE pressure induced in the liquid layer as calculated from Eq.~\eqref{GrindEQ__4_1_} for free boundaries, while for rigid  boundaries the values for free boundaries have to be multiplied by the Prandtl number-dependent factor $\left\langle \overline{F}\left(\widetilde{z}\right) \right\rangle _{z}$. We have obtained the required thermophysical-property information for these liquids from~\cite{RefProp}. For comparison we also present the values of $p_{{\rm EM}}$ calculated from Eq.~\eqref{GrindEQ__1_1_} and for $p_{{\rm c}}$ calculated from Eq.~\eqref{GrindEQ__1_2_} with a Casimir amplitude $\Delta=-0.15$ corresponding to periodic boundary conditions~\cite{Danchev}, which are conceptually closest to our case of stress-free boundary conditions. Indeed, the NE fluctuation-induced pressures are orders of magnitudes larger than those induced by critical fluctuations especially at $L=0.1$~mm and at $L=1$~mm, even for $\Delta{T}=-10$~K~\cite{miPRL2}. In Table~\ref{T1} we have presented the fluctuation induced NE pressures for $\Delta{T}=-25$~K, so that we can make a comparison below with thermophoretic experiments obtained with this temperature difference. The sign of the NE pressure is determined by the thermodynamic quantity between the square brackets in Eqs.~\eqref{GrindEQ__3_13_} and~\eqref{GrindEQ__3_17_}, which can be either positive or negative.  A review of the thermodynamic properties of a number of fluids indicates that the NE fluctuation induced pressure is positive for fluids in the liquid state and negative for fluids in the vapor state~\cite{RefProp,Perkins13}.

\begin{table*}[t]
\caption{Estimated Casimir pressures and NE fluctuation-induced pressures}
\begin{tabular*}{\textwidth}{c@{\extracolsep{\fill}}cccccc}
\toprule
& & &\multicolumn{2}{c}{$p_\text{NE}$\footnote{At $T=298$~K and $\Delta{T}=25$~K, for a pressure of 0.1~MPa}, free boundaries}
&\multicolumn{2}{c}{${p_\text{NE}}^{a,}$\footnote{For water $\text{Pr}\simeq 6.1$, for $n$-heptane $\text{Pr}\simeq 6.6$}, rigid boundaries}\\
$L$&$p_\text{EM}$\footnote{Eq.~(1)}&$p_\text{c}$\footnote{$p_\text{c}=-0.15 k_\text{B}T/L^3$}&Water&$n$-Heptane &Water&$n$-Heptane \\
\colrule
$10^{-6}$~m& $-1.3\times10^{-3}$~Pa  & $-6.1\times10^{-4}$~Pa&1.21~Pa&0.40~Pa &0.50~Pa&0.16~Pa\\
$10^{-4}$~m& $-1.3\times10^{-11}$~Pa & $-6.1\times10^{-10}$~Pa &$12.1\times10^{-3}$~Pa&$4.0\times10^{-3}$~Pa &$5.0\times10^{-3}$~Pa&$1.7\times10^{-3}$~Pa\\
\botrule
\end{tabular*}
\label{T1}
\end{table*}

Since our result for free boundaries is a rigorous one, while the one for rigid boundaries is an approximation, we focus in the remainder of this paper on NE pressures in the case of free boundaries. We need to emphasize that the NE pressure $p_\text{NE}(\widetilde{z};L)$, given by Eq.~\eqref{GrindEQ__3_12_}, not only depends on the thickness $L$ of the fluid layer, but also on the position $\widetilde{z}$ in the fluid layer. An important question is: what are the consequences of the tendency of the temperature gradient to induce a pressure gradient? To address this question, we first consider the case that there are no microparticles in the fluid. Then mechanical equilibrium would require
\begin{equation}
\frac{{dp}}{{d\widetilde{z}}} = \frac{{dp_{{\rm{LE}}} }}{{d\widetilde{z}}} + \frac{{dp_{{\rm{NE}}} }}{{d\widetilde{z}}} = 0.
\end{equation}
In the absence of any particles in the fluid layer, the only mechanism for the pressure gradient to relax is through a rearrangement of the density profile. We thus write
\begin{equation}
\rho(z) = \rho_\text{LE}(\widetilde{z}) + \rho _{{\rm{NE}}}(\widetilde{z};L),
\end{equation}
where $\rho _{{\rm{NE}}} \left( {\widetilde{z};L} \right)$ is a nonequilibrium contribution to the local-equilibrium density profile $\rho_\text{LE}(\widetilde{z})$, such that
\begin{equation}\label{E316B}
\rho _{{\rm{NE}}} \left( {\widetilde{z}};L \right) =  - \rho \kappa _T p_{{\rm{NE}}} \left( {\widetilde{z}};L \right).
\end{equation}
We note that the relationship~\eqref{E316B} is independent of the boundary conditions for the fluctuations.

More interesting is to consider the presence of microparticles in the fluid. Then we may compare the NE fluctuation induced forces with thermophoretic forces on microparticles in a temperature gradient which can be estimated from~\cite{SchermerEtAl}
\begin{equation} \label{GrindEQ__4_3_}
F_{{\rm th}} =-6\pi \eta RD_{{\rm th}} \nabla T,
\end{equation}
where $\eta$ is the dynamic viscosity of the liquid, $R$ the radius of the particle, and $D_{{\rm th}}$ its thermophoretic mobility in the liquid. For silica particles $D_{{\rm th}} =22\times 10^{-12} \; {\rm m}^{2} {\rm s}^{-1} {\rm K}^{-1}$ in water and $D_{{\rm th}} =5.7\times 10^{-12} \; {\rm m}^{2} {\rm s}^{-1} {\rm K}^{-1}$ in n-heptane as measured by Regazetti \emph{et al.}~\cite{RegazettiEtAl}. The experiments were performed on particles with a radius $R=3~\mu$m in a temperature gradient $\nabla T=25\; {\rm K}/0.1\; {\rm mm}$ in a liquid layer with $L=0.1$~mm. With $\eta=0.890\times10^{-3}$~Pa~s for water  and $\eta=0.387\times10^{-3}$~Pa~s for n-heptane, we find from Eq.~\eqref{GrindEQ__4_3_} $\left|F_{{\rm th}} \right|=0.28\times 10^{-12}$~N in water and $\left|F_{{\rm th}}\right|=0.03\times10^{-12}$~N in n-heptane. On the other hand, we find that the fluctuations induce a NE force, $F_{{\rm NE}}= \pi R^{2} p_{{\rm NE}}$, such that from the information in Table~\ref{T1}, $F_{{\rm NE}} \simeq 0.34\times 10^{-12}$~N in water and $F_{{\rm NE}} \simeq 0.12\times10^{-12}$~N in n-heptane.  However, these values are only averages and they vary strongly with the position $z$ in the liquid as can be seen from Eq.~\eqref{GrindEQ__1_5_} and Fig.~\ref{F1}. Thus in water $F_{{\rm NE}}$ will vary from 0 to $0.51\times 10^{-12}$~N in n-heptane from 0 to $0.18\times 10^{-12} \; {\rm N}$ inducing NE force differences over a height of $3\mu$m of about some tens of fN.  Hence, depending on the size of the particles, the magnitude of the temperature gradient, and of the choice of the liquid solvent, NE fluctuation-induced forces may become comparable to thermophoretic forces, and we may have discovered a new fluctuation induced thermophoretic force.

We note that Najafi and Golestanian~\cite{NajafiGolestanian} have considered an alternative nonequilibrium Casimir effect due to inhomogeneous noise correlations in a medium that is otherwise in local equilibrium. However, it has been demonstrated that nonequilibrium fluctuations arising from inhomogeneous noise correlations are orders of magnitude less significant than those arising from hydrodynamic couplings in the fluid by a temperature gradient \cite{miStatis}.

\section{EFFECT OF GRAVITY ON THE NE PRESSURE\label{S5}}

The NE temperature fluctuations are related to NE density fluctuations through Eq.~\eqref{GrindEQ__2_11_}. These thermal nonequilibrium fluctuations become so large that they are not ony affected by finite-size effects, but also by gravity. Kirkpatrick and Cohen \cite{KirkpatrickCohen2,KirkpatrickCohen} have pointed out that the structure factor in a fluid layer heated from below is strongly affected by gravity near the Rayleigh-B\'enard convective instability. Even more surprisingly, Segr\`e \textit{et al}. \cite{SegreSchmitzSengers} concluded that NE temperature fluctuations are strongly affected by gravity in fluid layers heated from above, \textit{i.e.,}  very far away from any hydrodynamic instability. These predictions have been confirmed by light-scattering~\cite{VailatiGiglio1} and shadowgraph experiments~\cite{TakacsEtAl2}. Hence, we should also expect that the NE pressure will be affected by gravity.

The combined finite-size effects and gravity on the NE temperature fluctuations in a bounded fluid layer have been analyzed by Ortiz de Z\'arate and Sengers \cite{Physica2,miPRE}. The effect of the gravitational field ${\bf g}$ enters through the Rayleigh number
\begin{equation} \label{GrindEQ__5_1_}
{\rm Ra}=\frac{\alpha L^{4}}{\nu D_{T} }~{\bf g}\cdot\boldsymbol{\nabla} T .
\end{equation}
The Rayleigh number is negative for fluid layers heated from above and positive for fluid layers heated from below. In this paper we only consider fluids in a stable quiescent state for which ${\rm Ra}\le {\rm Ra}_{{\rm c}}$, where ${\rm Ra}_{{\rm c}}$ is the critical Rayleigh number associated with the onset of thermal convection (for the case of free boundaries, $\text{Ra}_\text{c}=27\pi^4/4$).

In fluctuating hydrodynamics the temperature fluctuations can be calculated as a solution of two coupled differential equations, namely a fluctuating heat equation and a fluctuating Navier-Stokes equation \cite{BOOK}. If we follow again this procedure, but also include the gravitational contribution to the fluctuating Navier-Stokes equation, we find from Eq.~(27) in \cite{Physica2} that the expression~\eqref{GrindEQ__3_7_} for the NE structure factor $\widetilde{S}_{{\rm NE}}(k_{\parallel } ,z,z')$ changes into:
\begin{equation} \label{GrindEQ__5_2_}
\widetilde{S}_{{\rm NE}} \left(k_{\parallel } ,z,z'\right)=2L^{3} \sum _{1}^{\infty }\frac{k_{\parallel }^{2} L^{2} \sin \left(N\pi z/L\right)\sin \left(N\pi z'/L\right)}{\left(k_{\parallel }^{2} L^{2} +N^{2} \pi ^{2} \right)^{3} -{\rm Ra}{\kern 1pt} k_{\parallel }^{2} L^{2} }.
\end{equation}
Notice that, as expected, when $\text{Ra}=0$, Eq.~\eqref{GrindEQ__5_2_} reduces to Eq.~\eqref{GrindEQ__3_7_} above. If we substitute Eq.~\eqref{GrindEQ__5_2_} with $z'=z$ into Eq.~\eqref{GrindEQ__3_9_} and follow the same procedure as in Sec.~\ref{S3A}, we obtain
\begin{equation} \label{GrindEQ__5_3_}
\left\langle \delta T^2\left(z\right)\right\rangle _{{\rm NE}} =\frac{k_{{\rm B}} TL\left(\nabla T\right)^{2} }{48\pi \rho D_{T} \left(\nu +D_{T} \right)} F\left(\widetilde{z}{\rm ;Ra}\right)
\end{equation}
with
\begin{equation} \label{GrindEQ__5_4_}
F\left(\widetilde{z};{\rm Ra}\right)=48 \int _{0}^{\infty }\!\!d\widetilde{k} \sum _{1}^{\infty }\frac{\widetilde{k}^{3} \sin ^{2} \left(N\pi \widetilde{z}\right)}{\left(\widetilde{k}^{2} +N^{2} \pi ^{2} \right)^{3} -{\rm Ra}{\kern 1pt} \widetilde{k}^{2} }  .
\end{equation}
\begin{figure}[b]
\begin{center}
\includegraphics[width=0.5\columnwidth]{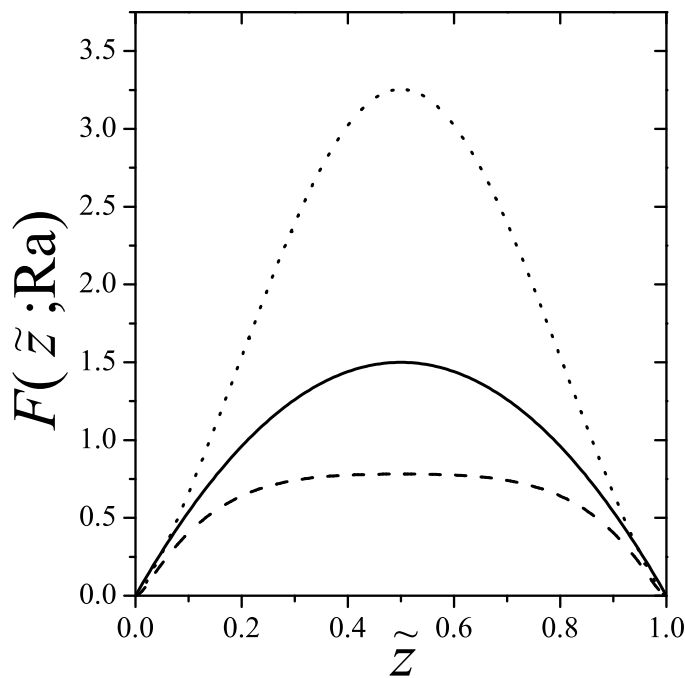}
\end{center}
\caption{Amplitude $F(\widetilde{z};{\rm Ra})$ of NE fluctuation-induced pressure for free boundaries, given by Eq.~\eqref{GrindEQ__5_4_}, as a function of $\widetilde{z}$ and for three values of the Rayleigh number. Dotted curve is for positive Ra=570, not far from the instability. Dashed curve is for large negative $\text{Ra}=-3000$. Solid curve is for zero Ra, in which case the amplitude is given by Eq.~\eqref{GrindEQ__3_11_}.}
\label{F2}
\end{figure}
Equation~\eqref{GrindEQ__5_3_} replaces Eq.~\eqref{GrindEQ__3_10_} with $F\left(\widetilde{z};0\right)=F\left(\widetilde{z}\right)$. In Fig.~\ref{F2} we show plots of the function $F\left(\widetilde{z};{\rm Ra}\right)$ for three values of the Rayleigh number. The function $F\left(\widetilde{z};{\rm Ra}\right)$ diverges when ${\rm Ra}$ approaches the critical value ${\rm Ra}_{{\rm c}}$ such that
\begin{align}
F\left(\widetilde{z};{\rm Ra}\right)& \to \frac{4\pi}{\sqrt{{\rm Ra}_{{\rm c}} -{\rm Ra}} } & \text{as}~{\rm Ra}&\to {\rm Ra}_{{\rm c}}.\label{GrindEQ__5_5_}
\end{align}
For negative ${\rm Ra}$, \textit{i.e.,} heating from above for regular fluids, the quiescent conductive state is always stable and its limiting behavior is given by
\begin{align}
F\left(\widetilde{z};{\rm Ra}\right)&\to \frac{6}{\left|{\rm Ra}\right|^{1/4} } -\frac{3\pi}{\sqrt{\left|{\rm Ra}\right|} } & \text{as}~{\rm Ra}&\to -\infty.\label{GrindEQ__5_6_}
\end{align}

If we substitute Eq.~\eqref{GrindEQ__5_3_} into Eq.~\eqref{GrindEQ__2_14_} and average again over the height $\widetilde{z}$ as in the derivation of Eq.~\eqref{GrindEQ__3_13_}, we now find for the average NE pressure:
\begin{equation} \label{GrindEQ__5_7_}
\begin{split}
p_{{\rm NE}} \left(L\right)&=\left\langle p_{{\rm NE}} \left(z;L\right)\right\rangle _{z}\\
&=\frac{c_{p} k_{{\rm B}} T^{2} \left(\gamma -1\right)}{96\pi D_{T} \left(\nu +D_{T} \right)} \left[1-\frac{1}{\alpha c_{p} } \left(\frac{\partial c_{p} }{\partial T} \right)_{p} +\frac{1}{\alpha ^{2} } \left(\frac{\partial \alpha }{\partial T} \right)_{p} \right]\langle F(\text{Ra})\rangle _{z} L\left(\frac{\nabla T}{T} \right)^{2}
\end{split}
\end{equation}
where the dependence on the Rayleigh number is contained in the average
\begin{equation} \label{GrindEQ__5_8_}
\langle F(\text{Ra})\rangle _{z} =\int _{0}^{1}d\widetilde{z} \, F\left(\widetilde{z};{\rm Ra}\right).
\end{equation}
We note that $\left\langle{F}(0)\right\rangle_z=1$, and Eq. \eqref{GrindEQ__5_7_} reduces to Eq. \eqref{GrindEQ__3_13_} in the absence of gravity.
\begin{figure*}[t]
\flushleft
\begin{center}
\includegraphics[width=0.45\textwidth]{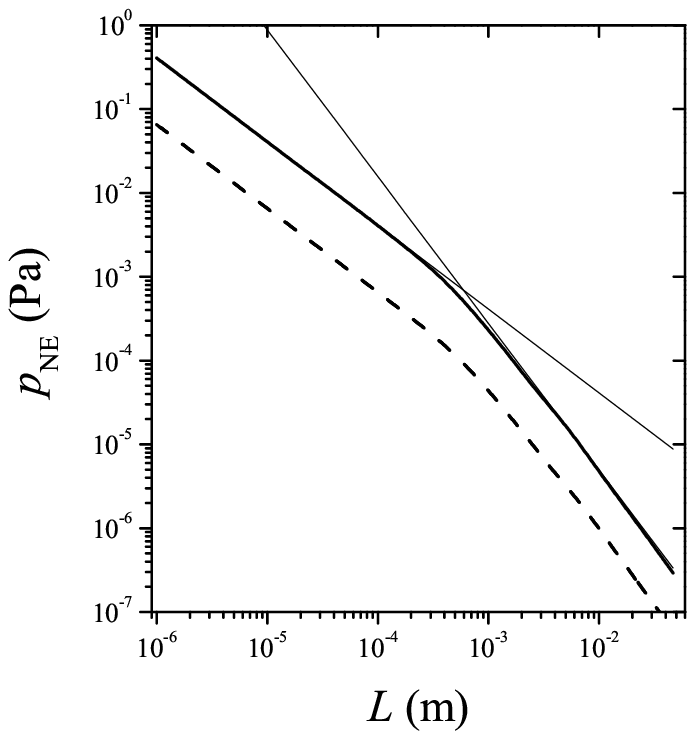}
\includegraphics[width=0.45\textwidth]{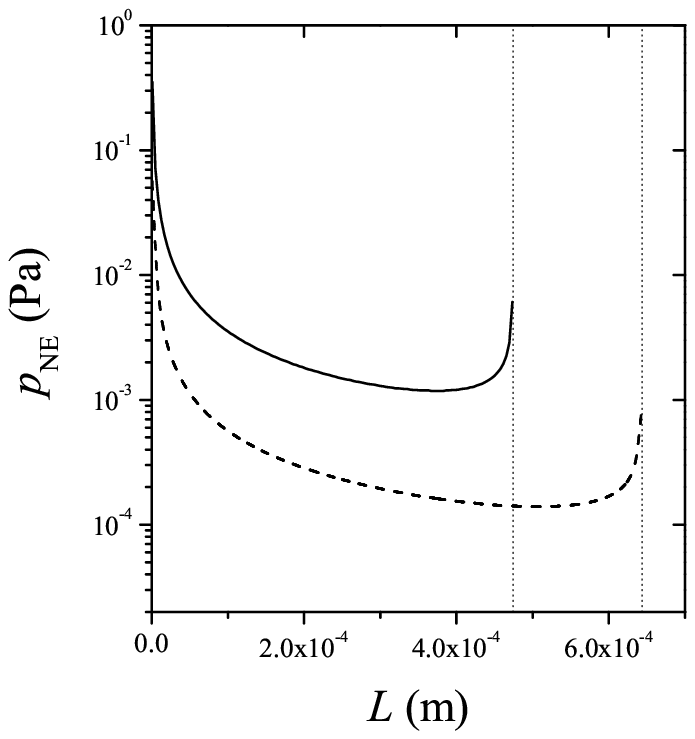}
\end{center}
\caption{Left panel: Double-log plot of the average NE pressure as a function of the separation $L$ between plates, for $n$-heptane at 25$^\circ$C when heated from above, for temperature differences of $\Delta{T}=-25$~K (solid thick curve) and $\Delta{T}=-10$~K (dashed curve). Thin straight lines indicate the two asymptotic behaviors for $\Delta{T}=-25$~K. Right panel: Logarithmic plot of the same average NE pressure as a function of $L$ when heated from below, for temperature differences of $\Delta{T}=25$~K (solid thick curve) and $\Delta{T}=10$~K (dashed curve). Thin dotted vertical lines indicate the plates separation $L_\text{c}$ at which the critical Rayleigh number is reached for each $\Delta{T}$.}
\label{F3}
\end{figure*}

In the left panel of Fig.~\ref{F3} we show, on a double-log scale, a plot of the average NE pressure, $p_{{\rm NE}}(L)$ given by Eq.~\eqref{GrindEQ__5_7_} for free boundaries, as a function of the plate distance $L$ when the fluid layed is heated from above (stable configuration). Thermophysical properties are for $n$-heptane at 25$^\circ$C. The solid curve is for a temperature difference of $\Delta{T}=-25$~K, and the dashed curve for a temperature difference of $\Delta{T}=-10$~K. One observes that the NE pressure exhibits a crossover form the $\propto \Delta{T}^2/L$ behavior for very small $L$ to a $\propto (|\Delta{T}|/L)^{7/4}$ behavior for large $L$. The latter can be easily obtained from Eq.~\eqref{GrindEQ__5_6_}. These asymptotic behaviors for the case of $\Delta{T}=-25$~K are indicated as thin straight lines in Fig.~\ref{F3}.

In the right panel of Fig.~\ref{F3} we show, on a logarithmic scale, a similar plot when the fluid layer is heated from below (unstable configuration). The solid curve is for a temperature difference of $\Delta{T}=25$~K, and the dashed curve for a temperature difference of $\Delta{T}=10$~K. Thin dotted vertical lines indicate the plates separation $L_\text{c}$ at which the critical Rayleigh number is reached for each $\Delta{T}$. We note that for positive Ra, essentially, the NE pressure exhibits a crossover form the $\propto \Delta{T}^2/L$ behavior for very small $L$ to a $\propto (\Delta{T})^{3/2}/\sqrt{L_\text{c}^3-L^3}$ behavior for $L$ close but below $L_\text{c}$.

From the information in Fig.~\ref{F3} we conclude that gravity has only a modest influence when the fluid layer is heated from above (negative $\Delta T$), but in fluid layers heated from below (positive $\Delta T$) the NE pressure diverges as the critical value of the Rayleigh number is approached.

\section{RELATIONSHIP WITH CORRELATION FUNCTIONS FOR LINEAR AND NON-LINEAR TRANSPORT COEFFICIENTS\label{S6}}

As we mentioned in the introduction, a NE pressure is related generally to a temperature gradient through a (non-linear) Burnett coefficient $\kappa _{{\rm NL}}$ defined in Eq.~\eqref{GrindEQ__1_5_}. The NE pressure is then caused by the fact that $\kappa _{{\rm NL}}$ diverges as $\kappa _{{\rm NL}}^{(1)} L$:
\begin{equation} \label{GrindEQ__6_1_}
p_{{\rm NE}} \left(L\right)=\kappa _{{\rm NL}}^{\left(1\right)} L\left(\nabla T\right)^{2} .
\end{equation}
Divergent Burnett and higher-order transport coefficients are not new and, in general, as the number of gradients increases, the strength of the divergence does as well~\cite{Brey,PomeauResibois,ErnstDorfman,KawasakiGunton,DuftyMcLennan,WongEtAl,Standish}. Hence, one would expect that Eq.~\eqref{GrindEQ__1_5_} or~\eqref{GrindEQ__6_1_} could be generalized into~\cite{SchepperEtAl}
\begin{equation} \label{GrindEQ__6_2_}
p_{{\rm NE}} \left(L\right)=\kappa^{\left(2\right)} \left(\nabla T\right)^{2} +\kappa ^{\left(4\right)} \left(\nabla T\right)^{4} +\kappa ^{\left(6\right)} \left(\nabla T\right)^{6} +...
\end{equation}
with $\kappa^{(2)}=\kappa _{{\rm NL}}^{(1)} L$. In Eq.~\eqref{GrindEQ__6_2_}, the coefficients $\kappa ^{\left(4\right)}$, $\kappa ^{\left(6\right)}$, are sometimes referred to as super Burnett coefficient, super-super Burnett coefficient, etc.~\cite{Foch,McLennan,DonevEtAl}\footnote{Examining the Fourier transformed hydrodynamic equations for fluctuations about a NESS with $\nabla{T}\neq0$ one finds that the nonequilibrium contributions $\sim\nabla{T}$ compete against equilibrium terms $\sim Dk^2$ with $D=D_T$ of $D=\nu$. The ratio of these two terms is $\sim\nabla{T}/Dk^2$. In infrared divergent integrals, $k^2\sim1/L^2$, so that in principle one would expect a formal expansion in powers of $L^2\nabla{T}$. Indeed, in the case of planar Couette flow such a formal expansion does occur}. In principle, the Burnett coefficients $\kappa ^{\left(n+2\right)}$ are expected to diverge as $\kappa^{(n+2)}\sim L^{1+2n}$, which would imply and even stronger NE pressure than the one proportional to $\left(\nabla T\right)^{2}$ in accordance with Eqs.~\eqref{GrindEQ__2_14_} or~\eqref{GrindEQ__6_1_}.  However, it can be shown that in the case of a temperature gradient all higher-order transport coefficients $\kappa ^{\left(n+2\right)}$ in Eq.~\eqref{GrindEQ__6_2_} for $n\ge 2$ have a zero prefactor. That is, Eq.~\eqref{GrindEQ__1_3_} for the NE structure factor and, hence, Eq.~\eqref{GrindEQ__2_14_} for the NE pressure are exactly proportional to $\left(\nabla T\right)^{2}$ and do not contain any higher-order contributions in the temperature gradient~\cite{KirkpatrickEtAl}. The fact that NE structure factor is proportional to $\left(\nabla T\right)^{2}$ has been extensively verified by light-scattering experiments~\cite{SegreEtAl1,Mixtures3}. It should be noted that the vanishing of higher-order transport coefficients generally does not occur in other NESS problems. For example, normal shear stresses in a steady-state uniform shear flow contain higher-order contributions in the velocity gradient~\cite{ErnstEtAl78}.

Our results also illustrate a number of other surprising features in nonequilibrium thermodynamics. The existence of a non-zero NE pressure or resulting NE density contribution represent a breakdown of the principle of local equilibrium,  a principle which has been considered one of the foundations of nonequilibrium thermodynamics~\cite{KestinDorfman}. On comparing Eq.~\eqref{GrindEQ__6_1_} with Eqs.~\eqref{GrindEQ__3_13_}, \eqref{GrindEQ__3_17_} and~\eqref{GrindEQ__5_5_}, we note that we have actually evaluated the non-linear Burnett coefficient $\kappa _{{\rm NL}}^{\left(1\right)}$. We find that the Burnett coefficient depends not only on the thermodynamic properties of the fluid, but also on boundary conditions and on gravity. We note that this is a general property of all transport coefficients including the common linear transport coefficients. The reason is that the hydrodynamic fluctuations, causing LTT contributions to the transport coefficients, depend on gravity which is a force in the momentum equation, and on the boundary conditions that affect the solutions of the fluctuating-hydrodynamics equations. The fact that transport coefficients of fluids depend on the boundary conditions was first pointed out by Nieuwoudt \textit{et al.}~\cite{NieuwoudtEtAl}. That long-range NE fluctuations cause the diffusion coefficient to become dependent on gravity was noticed by Brogioli and Vailati~\cite{VailatiBrogioli}. In many cases the dependence of the transport coefficients of fluids on boundary conditions and on gravity will be small, but not always. For instance, for a fluid layer heated from below, it has be shown~\cite{KirkpatrickCohen} that the dependence of the thermal conductivity of the fluid on gravity is so large that it actually diverges at the Rayleigh-B\'enard instability, just like the NE pressure does; in addition it also depends on the position in the fluid, like the Burnett coefficients. Note that the actual thermal conductivity of the fluid thus differs from the local-equilibrium value assumed for the thermal conductivity $\lambda$ in $D_{T} =\lambda /\rho c_{p}$ in the definition~\eqref{GrindEQ__5_1_} of the Rayleigh number.

In this paper we have restricted our analysis to NE pressures induced in a fluid by NE temperature fluctuations. One should expect similarly significant NE pressures induced in a fluid mixture by NE concentration fluctuations \cite{Mixtures3,VailatiGiglio1,GiglioNature,VailatiGiglio3,miPRL,miJCP,BrogioliEtAl2,VailatiEtAl}. This subject will be addressed in a subsequent publication.

\section{DISCUSSION\label{S7}}

Fluctuations in nonequilibrium steady states are always long range. In this paper we have shown that the long-range temperature fluctuations in a fluid subjected to a temperature gradient $\nabla T$ cause nonequilibrium (NE)  pressures $p_{{\rm NE}} \left(z;L\right)$ proportional to $\left(\nabla T\right)^{2}$ and depending on the location within the fluid. For a fixed value of the temperature gradient $\nabla T$ the NE pressures increase proportional to the distance $L$, while for a fixed value of the temperature difference $\Delta T$ the NE pressures vary as $L^{-1}$. As a consequence, these NE Casimir-like pressures are present over much larger distances $L$ than in the case of the Casimir forces thus far considered in the literature. The magnitude of the NE pressures does depend on the boundary conditions, as is the case for all other Casimir forces investigated thus far. However, the NE pressures also depend on gravity and for a fluid layer heated from below the NE pressure diverges as the Rayleigh-B\'enard instability is approached.

These fluctuation-induced phenomena imply a breakdown of the principle of local equilibrium which has been a fundamental principle in nonequilibrium thermodynamics. Traditionally, nonequilibrium thermodynamics is based on conservation laws supplemented with linear laws relating fluxes to forces with transport coefficients which were supposed to be material constants, independent of the solution of the hydrodynamic equations. This clear separation between conservation laws and linear material laws was violated when it was discovered that the correlation functions for the transport coefficients exhibit long-time tails leading to renormalization of the transport coefficients~\cite{BM74,BM74B}. However, what has not been appreciated is that transport coefficients also depend on gravity and on boundary conditions. These additional effects become significant in fluids heated from below. For instance, the thermal conductivity diverges at the Rayleigh-B\'enard instability. The divergence of the NE pressure and of the thermal conductivity near the Rayleigh-B\'enard instability is a dramatic demonstration of the breakdown of local equilibrium. Hence, long-range NE fluctuations make it necessary to rethink some fundamental features of nonequilibrium thermodynamics.

\section*{ACKNOWLEDGMENTS}

The research was supported by the U.S. National Science Foundation under Grant DMR-09-01907. We thank M.L. Huber, E.W. Lemmon, and R.A. Perkins of the U.S. National Institute of Standards and Technology for providing us with relevant thermodynamic-property information for the evaluation of the NE pressures. In addition, JOZ acknowledges support from the UCM/Santander Research Grant PR6/13-18867.

\bibliography{ortiz}

\end{document}